%% file: prevost_robust_2025.tex
\documentclass[journal]{IEEEtran}
\usepackage{graphicx}
\usepackage{cite}
\usepackage{picinpar}
\usepackage{amsmath}
\usepackage{url}
\usepackage{flushend}
\usepackage{colortbl}
\usepackage{soul}
\usepackage{multirow}
\usepackage{pifont}
\usepackage{amsfonts}
\usepackage{color}
\usepackage{alltt}
\usepackage[hidelinks]{hyperref}
\usepackage{enumerate}
\usepackage{breakurl}
\usepackage{epstopdf}
\usepackage{pbox}
\usepackage[printonlyused,nolist]{acronym}
\usepackage{booktabs}
\newcommand{\J}{\mathcal{J}}
\usepackage{amsthm}
\newtheorem*{assumption*}{\assumptionnumber}
\providecommand{\assumptionnumber}{}
\makeatletter

\newenvironment{assumption}[2]
 {%
  \renewcommand{\assumptionnumber}{Assumption #1$\mathcal{#2}$}%
  \begin{assumption*}%
  \protected@edef\@currentlabel{#1$\mathcal{#2}$}%
 }
 {%
  \end{assumption*}
 }

\newtheorem{theorem}{Theorem}
\newtheorem{remark}{Remark}

\begin{document}

\title{A robust mechanical sensorless control strategy for active rectification of small wind turbines}

\author{Adrien Pr\' evost, Vincent L\'echapp\'e, Romain Delpoux, Xavier Brun
  \thanks{Adrien Pr\'evost, Vincent L\'echapp\'e, Romain Delpoux and Xavier Brun are with Univ Lyon, INSA Lyon, Universit\'e Claude Bernard Lyon 1, Ecole Centrale de Lyon, CNRS, Amp\`ere, UMR5005, 69621 Villeurbanne, France
(e-mail: adrien.prevost@insa-lyon.fr; vincent.lechappe@insa-lyon.fr; romain.delpoux@insa-lyon.fr; xavier.brun@insa-lyon.fr)}
      
}

\maketitle
	
\begin{abstract}
This article proposes a mechanical sensorless control strategy for the synchronous rectification of small wind turbines equipped with a surface-mounted \ac{PMSG}. By means of Lyapunov theory, the \ac{GAS} of the closed loop system is proven. It allows the use of a classical Sliding Mode Observer (SMO) to remove the mechanical sensor in the control loop despite uncertainties on the resistance and inductance parameters. The analysis of the equilibrium points have made it possible to propose an analytic model of the angular misalignment between the true and the observer rotating frames, responsible for current tracking errors. Experimental tests on a wind turbine emulator show that despite large errors on the the resistance and inductance parameters, the impact on the energy harvest is low, proving that the strategy's performance is robust to high uncertainties.

\end{abstract}

\begin{IEEEkeywords}
\ac{WECS}, \ac{PMSG}, robustness, mechanical sensorless control, sliding mode observer, synchronous rectifier
\end{IEEEkeywords}


\definecolor{limegreen}{rgb}{0.2, 0.8, 0.2}
\definecolor{forestgreen}{rgb}{0.13, 0.55, 0.13}
\definecolor{greenhtml}{rgb}{0.0, 0.5, 0.0}

\input{a_acronyms}

\section{Introduction}

\IEEEPARstart{S}{ynchronous} rectification can increase energy production of \ac{PMSG} based small wind turbines in comparison with the most widespread diode rectifier solution \cite{mirecki_architecture_2007, 
prevost_experimental_2023}. 
This energy conversion strategy requires the knowledge of the rotor electrical angular position to apply \ac{FOC}. It is classically obtained with a position sensor which can be source of failures and adds extra design costs. Position sensorless \ac{FOC} allows the implementation of fully mechanical sensorless wind turbine control laws (i.e.\ without position nor anemometer sensor), a domain of high interest in the literature \cite{
brahmi_comparative_2009, qiao_wind_2012, 
buticchi_active_2015, aubree_design_2016
}. A \acf{BEMF} observer is typically employed to reconstruct the electrical angle and velocity. This type of observer is model-based and requires the knowledge of the electrical parameters, namely \ac{PM} flux linkage, inductance and resistance. In the area of small wind turbines, these parameters are not always well known and may vary for each wind turbine site. 
The generator manufacturing process can also be source of parameter uncertainty, in particular when the machines are hand-crafted \cite{latoufis_axial_2012}. Moreover, some elements are specific, such as the length and section of the transmission cable. It implies that for an identical generator, there could be different set of parameters. Finally, some parameters can also vary during operation. Typically, the winding resistance and flux linkage are affected by the operating temperature and the inductance can saturate. 

The use of model-based observers with wrong electrical parameters unavoidably leads to angle estimation errors \cite{batzel_electric_2005, bolognani_design_2014, seilmeier_impact_2014, wang_position_2020-1} 
which has given rise to several  works. For instance in \cite{batzel_electric_2005}, angle observation errors due to wrong inductance and resistance parameters were quantified for a \ac{BEMF} linear extended observer. 
In \cite{lu_artificial_2013}, a method to reduce the observation error due to inductance saturation was proposed based on the concept of artificial inductance. In \cite{zhao_effective_2023}, an effective position error compensation method based on a disturbance observer is employed and shows very good results for variations of inductance. These methods address the problem of parameters variation during operation and requires a good knowledge of the q-axis inductance, but they are not suited for the parameter uncertainty issue. In \cite{benevieri_experimental_2022}, five amongst the most promising \ac{BEMF} observers for surface mounted synchronous PM machines including a robust adaptive algorithm were compared experimentally. Tests with a 50\% error on the inductance value resulted in significant position estimation errors for all observers. Another strategy rely on parameter identification. It can be performed online \cite{zhu_online_2021} and without mechanical sensor \cite{delpoux_parameter_2014}. To obtain the real system parameters, the identification should be conducted in situ, without position sensor and embedded in the final hardware (e.g.\ microcontroller), which adds extra software complexity. Moreover, some identification methods require specific experimental setups which are not compatible with wind turbine application (e.g.\ shaft speed being imposed by another motor \cite{armando_experimental_2013}). 

Finally, even if a parameter identification method is embedded in the wind turbine system, obtaining the exact machine parameters over the whole speed and temperature ranges remains a challenge. For the reasons mentioned above, the ideal case where the system parameters are perfectly known is unrealistic. 

The goal of this article is to propose a robust mechanical sensorless strategy for small wind turbines with a non-salient \ac{PMSG} which electrical parameters are highly uncertain. In this scope, our first contribution is the finding of a controller gain condition implying that the system is \ac{GAS}. Based on an existing \ac{SMO}, the second contribution consists in proposing gain tuning conditions of the observer to ensure its stability despite uncertainties on the resistance and inductance. The third contribution is the derivation of an analytical model to compute the influence of parameter errors on the generator operating points. Although studied in the case of an \ac{AFPM} generator, the presented results are valid for all Surface mounted Permanent Magnet Synchronous Machines. Finally, in terms of energy harvesting, it is shown experimentally that the performance of the sensorless strategy is robust to large parameter uncertainties.
\color{black}
The paper is organized as follows. The \ac{WECS} is modelled in Section \ref{sec:modelling}. The WECS control scheme is described in Section \ref{sec:wecs} and the observer is designed in Section \ref{sec:obs}. The robustness analysis of the closed-loop system is conducted in Section \ref{sec:robustness}. In Section \ref{sec:results}, the effectiveness of the proposed strategy is demonstrated experimentally and discussed.

\section{Modelling of the WECS} \label{sec:modelling}

\begin{figure}[h]
\begin{center}
\includegraphics[width=0.37\textwidth]{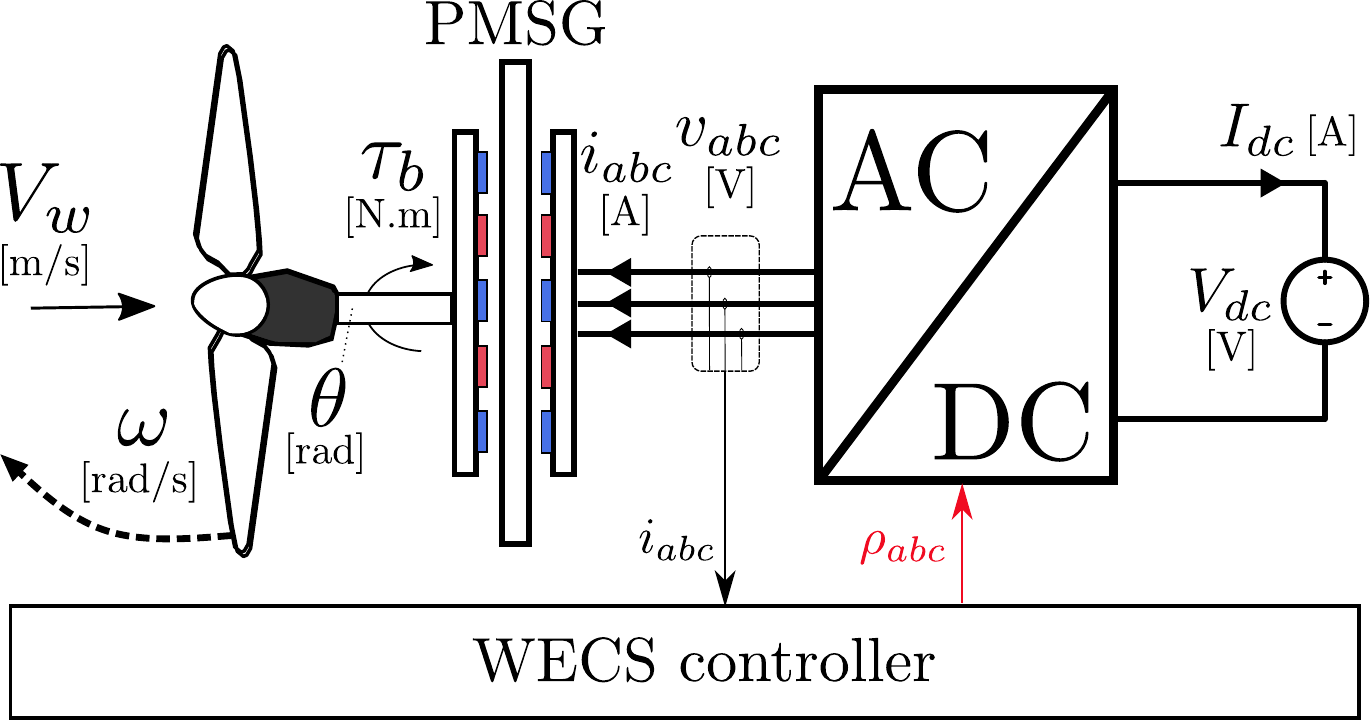}
\caption{The \acf{WECS} architecture.}
\label{fig:wecs_overview}
\end{center}
\end{figure}

A schematic of the \ac{WECS} architecture is given on Fig. \ref{fig:wecs_overview}. 

\subsection{Aerodynamics}

The aerodynamic efficiency of the wind turbine rotor is commonly modelled with the power coefficient $C_p$ which enables to write an expression of the aerodynamic power extracted from the wind by the blades

\begin{equation}\label{eq:p_m}
P=\frac{1}{2}\rho_{air} AV_{w}^3C_p(\lambda),
\end{equation} with A the rotor swept area, $\rho_{air}$ the air density and $V_w$ the wind velocity.
 
The case-study wind turbine blades have a fixed pitch, thus $C_p$ varies in function of the \ac{TSR} $\lambda$ written as
\begin{equation}\label{eq:tsr}
\lambda=\frac{\omega\cdot R_r}{V_w}
\end{equation}
where $\omega$ is the rotor angular velocity and $R_r$ the rotor radius. By conservation of power, the aerodynamic torque applied by the blades on the shaft can be written as 
\begin{equation}\label{eq:taub}
\tau_b=\frac{P}{\omega}.
\end{equation}

\subsection{Rotating system}
The wind turbine is direct-drive, hence the rotor dynamics can be expressed as a single inertia equivalent system given by 
\begin{equation}\label{eq:efd_eol}
J\frac{d\omega}{dt}=\tau_b+\tau_g-b\omega
\end{equation}
with $\tau_g$ the electromagnetic torque applied by the \ac{PMSG}, $J$ and $b$ the total rotating system inertia and damping coefficient respectively.

\subsection{Generator}
Here, we consider a non-salient \ac{PMSG}. The corresponding current dynamics in motor convention are given by
\begin{equation}
L\frac{di_{abc}}{dt} = v_{abc}-Ri_{abc}- e_{abc}
\end{equation}
with $R$ the phase resistance, $L$ the phase inductance, $v_{abc}=[v_a\quad v_b\quad v_c]^T$ the phases voltages, $i_{abc}=[i_a\quad i_b\quad i_c]^T$ the phase currents and $e_{abc}=[e_a\quad e_b\quad e_c]^T$ the \acp{BEMF}. The latter are induced by the \acp{PM} according to
\begin{equation}
e_{abc}=\frac{d}{dt}\lambda_r.
\end{equation} 
The PM flux linkage component $\lambda_r$ is given by
\begin{equation}
\lambda_r = \phi_f \left[ \begin{matrix}
\cos{\left(\theta_e\right)} \\
\cos{\left(\theta_e-\frac{2\pi}{3}\right)} \\
\cos{\left(\theta_e+\frac{2\pi}{3}\right)} \\
\end{matrix} \right]
\end{equation}
with $\phi_f$ the peak magnetic flux of the \acp{PM} seen by a stator phase winding and $\theta_e$ the rotor angular electrical position. 

Through Clarke transformation, it is possible to write the generator currents' dynamics in the $\alpha-\beta$ frame such as
\begin{equation}
\begin{array}{ll}\label{eq:diab}
\dfrac{di_{\alpha\beta}}{dt}=\dfrac{1}{L}v_{\alpha\beta}-\dfrac{R}{L}i_{\alpha\beta}-\dfrac{1}{L}e_{\alpha\beta},
\end{array}
\end{equation}
with the BEMF in the $\alpha-\beta$ frame given by
\begin{equation}\label{eq:eab}
\begin{array}{ll}
e_{\alpha\beta}=p \phi_f\omega \begin{bmatrix}
-\sin{(\theta_e)}\\\cos{(\theta_e)}
\end{bmatrix}.
\end{array}
\end{equation}
Park transformation 
allows describing the machine behavior in the $d-q$ rotating frame which is conveniently used for \ac{FOC}. Its construction usually requires the knowledge of the rotor angular electrical position since it is defined as
\begin{equation}\label{eq:park}
\mathcal{P}(\theta_e)=\begin{bmatrix}
			\cos(\theta_e) & \sin(\theta_e)\\
			-\sin(\theta_e) & \cos(\theta_e)\\
		\end{bmatrix}
\end{equation}
with $x_{dq}=\mathcal{P}(\theta_e)x_{\alpha\beta}$ expressing the projection of the variable $x$ from $\alpha - \beta$ to $d-q$ frame.
In the case of a non-salient \ac{PMSG}, the currents' dynamics in the $d-q$ frame are given by
\begin{equation} \label{eq:vdq}
L\frac{di_{dq}}{dt}=v_{dq}-Ri_{dq}-p\omega L\begin{bmatrix}0& -1 \\1& 0\end{bmatrix}\ i_{dq}-p\phi_f\omega \begin{bmatrix}0\\1\end{bmatrix}
\end{equation}
with $p$ the pole pairs number.
The electromagnetic torque of the \ac{PMSG} is proportional to the quadrature axis current according to
\begin{equation}\label{eq:tau_g}
\tau_g=\frac{3}{2}p\phi_fi_q.
\end{equation}


\begin{remark}
    The rectifier used in the article a classical three legs with two levels voltage source converter structure. The electrical power at the rectifier output is given by $P_{dc}=V_{dc}\cdot I_{dc}$.
\end{remark} 

\section{WECS control scheme}\label{sec:wecs}

\subsection{Problem statement}
Let $R_o$, $L_o$ be the assumed parameters values used to design the WECS control law such as
\begin{equation}
  R_o = R+\delta_R, ~
  L_o = L+\delta_L
\end{equation}
with $\delta_R$, $\delta_L$ the uncertainties on the resistance and inductance respectively. The goal is to propose a WECS control law that maximizes the wind turbine energy capture while guarantying the system stability, in spite of the uncertainties $\delta_R$, $\delta_L$. The paper will focus on the maximum power operation zone of the WECS, also called "zone 2" \cite{johnson_control_2006}. Moreover, the following assumption will be used for the design of the control law.

\begin{assumption}{1}{}\label{a:bounds}
The parameter uncertainties are bounded such as 
\begin{equation}
L_o,L \in [L_{min};L_{max}]\text{ and }
R_o,R \in [R_{min};R_{max}]
\end{equation}
with $L_{min}>0$, $L_{max}$, $R_{min}>0$, $R_{max}$ defined as the assumed minimum and maximum bounds of the inductance and resistance respectively.
\end{assumption}

\subsection{Optimal Torque Control}
\ac{OTC} is a simple wind turbine control strategy which does not need information on the wind velocity.
 In spite of its sensitivity to high inertia, it can lead to the optimal power in steady state while ensuring smooth shaft loads during transients. This makes \ac{OTC} particularly adapted for \acp{SWT} which have a low rotor inertia.
The original \ac{OTC} strategy consists in controlling the generator torque according to 
\begin{equation}\label{eq:OT}
\tau_g^\#=-K\omega^2,
\end{equation}
where $K$ is the torque gain. 
The torque leading to the maximal aerodynamic power extraction is called optimal torque. The latter can be obtained by setting the torque gain as:
\begin{equation}\label{eq:kopt_exp}
K_{opt}=\frac{1}{2}\rho_{air}AR_r^3\frac{C_{p(max)}}{\lambda_{(opt)}^3}
\end{equation}
with $C_{p(max)}$ the maximum power coefficient which occurs at the optimal \ac{TSR} $\lambda_{(opt)}$. 
Eqs. \eqref{eq:tau_g} and \eqref{eq:OT} allow the computation of the quadrature current reference $i_{q}^\#$ while due to surface mounted \ac{PM}, the minimization of joules losses, without flux weakening implies $i_{d}^\# = 0$. One can deduce the following generator currents set points 
\begin{equation}\label{eq:idqref}
i_{dq}^\#=\begin{bmatrix}
0\\ -\dfrac{2K_{opt}\omega^2}{3p\phi_f}
\end{bmatrix}
\end{equation}
leading to the optimal torque which can be tracked through \ac{FOC}. As no mechanical sensor is used in this work, the turbine rotational velocity will be reconstructed with an observer (see Section \ref{sec:obs}). Hence, in Eq. \eqref{eq:idqref} the speed estimate $\hat \omega$ replaces $\omega$.

\subsection{Field Oriented Control}

\begin{figure}[h]
\begin{center}
\includegraphics[width=0.37\textwidth]{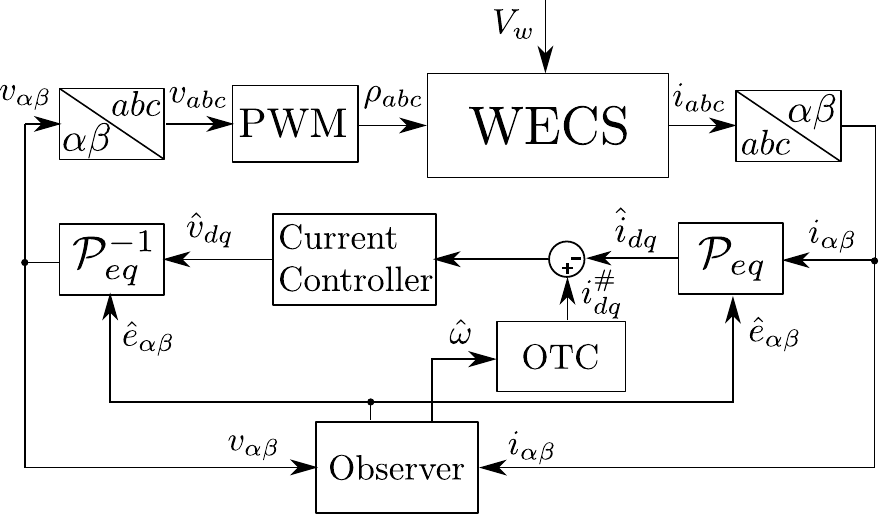}
\caption{WECS control scheme overview.}
\label{fig:control_scheme}
\end{center}
\end{figure}

To control the generator torque required by the \ac{OTC} algorithm, the well-established \ac{FOC} method is employed. As aforementioned, the computation of Park matrix $\mathcal{P}(\theta_e)$ requires the electrical angular position $\theta_e$ of the machine. It is well known that the observation of $e_{\alpha\beta}$ makes it possible to compute an equivalent of Park matrix given by 
\begin{equation}
 \mathcal{P}_{eq}=\dfrac{1}{\sqrt{{\hat e_\alpha}^2+{\hat e_\beta}^2}}\begin{bmatrix}
			{\hat e_\beta} & -{\hat e_\alpha}\\
			{\hat e_\alpha} & {\hat e_\beta}\\
		\end{bmatrix}
\end{equation} with $\hat e_{\alpha\beta}$ the BEMF estimates (see Section \ref{sec:obs}). This allows the projection of variables in the estimated Park frame that we call $\hat d-\hat q$, such as $\hat{x}_{dq}=\mathcal{P}_{eq}x_{\alpha\beta}$.
It should be noted that since the sine waves amplitudes are normalized, the transformation is not sensitive to any error made on the observed BEMFs amplitude. For position sensorless \ac{FOC}, the regulation of the currents has to happen in the $\hat d-\hat q$ frame obtained through $\mathcal{P}_{eq}$. An overview of the \ac{WECS} control scheme used in this article is given on Fig. \ref{fig:control_scheme}. A current controller is employed to enable the OTC strategy, it is described in the next section. Moreover, an observer is needed to reconstruct the rotor velocity and \acp{BEMF}, which will be addressed in the section \ref{sec:obs}.

\section{Current controller design}

In absence of parameter uncertainty, the system described by eqs. \eqref{eq:efd_eol} and \eqref{eq:vdq} results in
\begin{equation}
    \left\lbrace
\begin{array}{lll}
\dfrac{di_{d}}{dt}&=&\dfrac{1}{L} v_{d}-\dfrac{R}{L}i_{d}+p\omega i_q,\\
&&\\
\dfrac{di_{q}}{dt}&=&\dfrac{1}{L} v_{q}-\dfrac{R}{L}i_{q}-p\omega i_d-\dfrac{p\phi_f}{L} \omega,\\
&&\\
\dfrac{d\omega}{dt}&=&\dfrac{1}{J}\tau_b+\dfrac{3p\phi_f}{2J} i_q-\dfrac{b}{J}\omega.\\
\end{array}
\right.
\end{equation}
The currents controllers are synthesized with a state feedback and an integral action such as
\begin{equation}\label{eq:ctrl_d}
    \left\lbrace
\begin{array}{lll}
v_d &=&-k_pi_{d}-k_i x_{id}\\
\dot{x}_{id} &=& e_{d}\\
\end{array}
\right.
\end{equation}
with $e_d  =i_d-i^{\#}_{d}$ and
\begin{equation}\label{eq:ctrl_q}
    \left\lbrace
\begin{array}{lll}
v_q &=&-k_pi_{q}-k_i x_{iq}\\
\dot{x}_{iq} &=&e_q\\
\end{array}
\right.
\end{equation}
with $e_q=i_q-i^{\#}_{q}$. 

\begin{remark}
    Here, no linearization is applied in the control scheme, the system is kept nonlinear.
\end{remark}

For given constant blades torque and current references, let also be $e_{\omega}=\omega-\omega^{*}$ with $\omega^{*}=\frac{3}{2b}p\phi_fi^{\#}_q+\frac{\tau_b}{b}$. We additionally define 
\begin{equation}
    \left\lbrace
\begin{array}{lll}
e_{i_d}&=&x_{i_d}-x_{i_d^{*}},\\
e_{i_q}&=&x_{i_q}-x_{i_q^{*}},\\
\end{array}
\right.
\end{equation}
with the closed-loop equilibrium states in the $d$ and $q$ axis given by
\begin{equation}
    \left\lbrace
\begin{array}{lll}
x_{i_d^{*}}&=&\dfrac{-(k_p+R)i^{\#}_d+pL\omega^{*} i^{\#}_q}{k_i},\\
x_{i_q^{*}}&=&\dfrac{-(k_p+R)i^{\#}_q-pL\omega^{*} i^{\#}_d-p\phi_f\omega^{*}}{k_i}.\\
\end{array}
\right.
\end{equation}
Given these definitions, eqs. \eqref{eq:ctrl_d} and \eqref{eq:ctrl_q} can be re-written as
\begin{equation}
    \left\lbrace
\begin{array}{lll}
v_d &=&-k_p (e_{d}+i^{\#}_{d})-k_i (e_{id}+x_{i_d^{*}}),  \\
\dot{x}_{id} &=& e_{d}\\
\end{array}
\right.
\end{equation}
for the $d$ axis, and 
\begin{equation}
    \left\lbrace
\begin{array}{lll}
v_q &=&-k_p(e_{q}+i^{\#}_{q})-k_i (e_{iq}+x_{i_q^{*}})\\
\dot{x}_{iq} &=&e_q\\
\end{array}
\right.
\end{equation}
for the $q$ axis.

Hence the closed loop system dynamics can be given by
\begin{equation}\label{eq:cl_system}
\left\lbrace
\begin{array}{lll}
\dot{e}_d &=& \dfrac{-(k_p+R){e_d}-k_i e_{i_d}}{L} \\ && +p(e_{\omega}+\omega^{*}) {e_q}+pe_{\omega} {i^{\#}_{q}} \\
&&\\
\dot{e}_q  &=&  \dfrac{-(k_p+R){e_q}-k_ie_{i_q}{-p\phi_fe_{\omega}}}{L} \\ && -p(e_{\omega}+\omega^{*}) {e_d}-pe_{\omega} {i^{\#}_d}\\
\dot e_{i_d}&=&e_d \\
\dot{e}_{i_q} &=&e_q \\
\dot e_{\omega}&=&\dfrac{\frac{3}{2}p\phi_f}{J} {e_q}-\dfrac{b}{J}e_{\omega}\\
\end{array}
\right.
\end{equation}

\begin{theorem}\label{t:syst_stability}
The closed-loop system given by eq. \eqref{eq:cl_system} is \ac{GAS} if the following condition is met
\begin{equation}\label{eq:kp}
\left\lbrace
\begin{array}{lll}
        k_p&>&a-R\ \\ k_i&>&0 
\end{array}        
        \right.
\end{equation}
with $a=\dfrac{3}{4b}p\phi_f\sqrt{p^2(\phi_f+Li_d^\#)^2+(L{i_q^\#}p)^2}-\dfrac{3p^2\phi_f(\phi_f+Li_d^\#)}{4b}$.
\end{theorem}
The proof is detailed in the Appendix.
\begin{remark}
In Ortega et al. \cite{ortega_permanent_2018}, a similar result is given for a classical PI controller structure and a current reference in the d axis set to zero ($i_d^\#=0$). Our result is different and complementary since the controller is a state feedback with integral action and the reference $i_d^\#$ does not need to be zero.
\end{remark}
If $i_d^\#=0$, as in eq. \eqref{eq:idqref}, the expression of $a$ can be simplified
    \begin{equation}\label{eq:a}
       a=-\frac{3\,p\,\phi_f \,{\left(p\,\phi_f -\sqrt{p^2 \,{\left(L^2 \,{i_q^\# }^2 +{\phi_f }^2 \right)}}\right)}}{4\,b}
    \end{equation}
From eqs. \eqref{eq:kp} and  \eqref{eq:a}, the minimum value of the gain $k_p$ to guarantee the system stability can be computed according to the system parameters ($\phi_f, p, R, L, b $) and the current set-point $i_q^\#$. To ensure stability in the whole operating range, $k_p$ should be calculated considering the maximum value of the current reference. One can deduce that the gain $k_p$ is likely to be high if the maximum current reference is high and the friction coefficient is low.

\begin{remark}
    In presence of uncertainties on the parameters $R, L$, the assumed values $R_{min}$ and $L_{max}$ shall be used to calculate $k_p$.
\end{remark}

\begin{remark}
    Theorem 1 does not cover the stability of the \ac{OTC} law, which was already addressed by Johnson et al. \cite{johnson_control_2006}.
\end{remark}

\color{black}

\section{Observer design}\label{sec:obs}

\subsection{Currents observer design}
The goal here is to design an observer which can lead to the reconstruction of the equivalent Park transformation $\mathcal{P}_{eq}$ in spite of uncertainties on the electrical parameters $R$ and $L$.

Define the following current \ac{SMO} \cite[p. 266]{utkin_sliding_2009} as 
\begin{equation}\label{eq:diabhat}
\frac{d\hat i_{\alpha\beta}}{dt} = \frac{1}{L_o}v_{\alpha\beta}-\frac{R_o}{L_o}\hat i_{\alpha\beta}-\frac{l_1}{L_o}\text{sign} (\hat i_{\alpha\beta} - i_{\alpha\beta})
\end{equation}
where $l_1$ is the sliding mode gain. 
Subtraction of Eq. \eqref{eq:diab} to \eqref{eq:diabhat} gives rise to the sliding surface $ s_{\alpha\beta} = \hat i_{\alpha\beta} - i_{\alpha\beta}$ which mismatch dynamics are given by
\begin{equation}
\frac{ds_{\alpha\beta}}{dt} = \frac{1}{L}e_{\alpha\beta} +\frac{1}{L_o}\Pi_{\alpha\beta} -\frac{R_o}{L_o} s_{\alpha\beta} -\frac{l_1}{L_o}\text{sign} (\bar s_{\alpha\beta}).
\end{equation}
Here, the term $\Pi_{\alpha\beta}$ contains the perturbations due to parameter uncertainties according to
\begin{equation}
\Pi_{\alpha\beta}=\frac{R\delta_L-\delta_RL}{L}i_{\alpha\beta}-\frac{\delta_L}{L} v_{\alpha\beta}.
\end{equation}

When the parameters are perfectly known meaning that $\delta_L=0$ and $\delta_R=0$, the term $\Pi_{\alpha\beta}$ is equal to zero. In this academic case, the condition on the gain $ l_1>\text{max}(\mid e_\alpha\mid,\mid e_\beta\mid) $ implies that $s_{\alpha\beta}\dot s_{\alpha\beta}<0$, leading to sliding mode after a finite time interval. However, in the presence of uncertainties on $L$ and $R$, this condition should be revised.
Indeed, when $\Pi_{\alpha\beta}\neq 0$ the condition $s_{\alpha\beta}\dot s_{\alpha\beta}<0$ necessary to sliding mode implies that
\begin{equation}\label{eq:l1_theo}
l_1>\text{max}\left(\left | \dfrac{L_o}{L}e_\alpha+\Pi_{\alpha} \right |, \left | \dfrac{L_o}{L}e_\beta+\Pi_{\beta} \right | \right).
\end{equation} 
In practice, condition described by Eq. \eqref{eq:l1_theo} is not straight forward to use as is. Hence, we propose another condition on $l_1$ which is function of the parameters bounds and system extreme operating points.
Let us define the maximum $\alpha - \beta $ frame axis \ac{BEMF}, phase voltage and current amplitudes  $E_{max}$, $V_{max}$ and $I_{max}$ such as 
\begin{equation}
\begin{array}{lcl}
E_{max}&=& \text{max}(\mid e_\alpha\mid,\mid e_\beta\mid),\\
V_{max}&=& \text{max}(\mid v_\alpha\mid,\mid v_\beta\mid),\\
I_{max}&=& \text{max}(\mid i_\alpha\mid,\mid i_\beta\mid).
\end{array}
\end{equation}
The following condition guarantees that $s_{\alpha\beta}\dot s_{\alpha\beta}<0$ provided that Assumption \ref{a:bounds} holds (see proof in Appendix)

\begin{equation}\label{eq:l1_robust}
l_1>\dfrac{L_o}{L_{min}}E_{max}+\left( R_{max}\dfrac{\Delta_L}{L_{min}}+\Delta_R\right)I_{max}+\dfrac{\Delta_L}{L_{min}}V_{max}
\end{equation}
with $\Delta_R=R_{max}-R_{min}>0$ and $\Delta_L=L_{max}-L_{min}>0$.

The computation of $l_1$ requires a knowledge of the extreme operating points $E_{max}$, $V_{max}$ and $I_{max}$ of the system.
However, this necessitates to find the equilibrium points of the closed loop system in presence of parameter uncertainties. This could be achieved by modelling the \ac{BEMF} observation error due to these uncertainties. The next subsection will detail the BEMF observer. Following this, a modelling of the observation error will be proposed in Section \ref{sec:robustness}.

\subsection{BEMF and speed observer}

Assume that $l_1$ is chosen so that the condition $s_{\alpha\beta}\dot s_{\alpha\beta}<0$ is satisfied, then sliding mode occurs after a finite time interval. Once the sliding surface is reached, $s_{\alpha\beta}= \dot s_{\alpha\beta}=0$ which results in the chattered signal
\begin{equation}
z_{\alpha\beta}=l_1 \text{sign} (\bar s_{\alpha\beta})=\frac{L_o}{L}e_{\alpha\beta}+\Pi_{\alpha\beta}+\Pi_{HF}
\end{equation}
with $\Pi_{HF}$ the perturbation due to the high frequency switching action. To obtain the smooth estimates $\hat e_{\alpha\beta}$, the $z_{\alpha\beta}$ signals needs to be low-pass filtered. The observer detailed in \cite[p. 267]{utkin_sliding_2009} has low-pass filtering properties and the advantage of providing an estimate $\hat{\omega}$ of the rotational velocity. We propose a slightly modified version of this observer by inserting the gain $l_3$ such as
\begin{equation}\label{eq:obs_emf} \left\{
\begin{array}{lcl}
\dfrac{d\hat {e}_\alpha}{dt}  &=& - \hat \omega_e \hat e_\beta -l_2({\hat e}_\alpha-z_\alpha),\\
\dfrac{d\hat {e}_\beta}{dt} &=& \hat \omega_e \hat e_\alpha -l_2({\hat e}_\beta-z_\beta),\\
\dfrac{d\hat \omega_e}{dt}  &=& l_3\left[(\hat e_\alpha-z_\alpha)\hat e_\beta - (\hat e_\beta-z_\beta)\hat e_\alpha\right] ,
\end{array} \right.
\end{equation}
where $\omega_e=p\omega$ is the electrical rotational velocity and $l_2>0$, $l_3>0$ are constant observer gains.
The convergence of the observer is proven for $l_2>0$, $l_3=1$ and without parameter uncertainty \cite[p. 267]{utkin_sliding_2009}. Choosing $l_3>1$ has shown to accelerate the convergence of $\hat{\omega}_e$ to $\omega_e$. The convergence proof for the general case is out of the scope of this article.

\section{Robustness analysis}\label{sec:robustness}

\subsection{Modelling of the BEMF observation error}

Thanks to the low-pass filter property of Eq. \eqref{eq:obs_emf}, the high frequency $\Pi_{HF}$ are filtered, and one obtains
\begin{equation}\label{eq:ehat_1}
\hat e_{\alpha\beta}=\frac{L_o}{L}e_{\alpha\beta}+ \frac{R\delta_L-\delta_RL}{L}\mathcal{P}^{-1}(\theta_e)i_{dq}-\frac{\delta_L}{L}\mathcal{P}^{-1}(\theta_e)v_{dq}
\end{equation}
at the observer output. Rewriting \eqref{eq:vdq} at steady state gives the expression of $v_{dq}$ at equilibrium given by
\begin{equation}\label{eq:vdq_std}
v_{dq}^*=Ri_{dq}^*+p\omega L\J i_{dq}^*+e_{dq}^*.
\end{equation}
Substituting Eqs. \eqref{eq:eab} and \eqref{eq:vdq_std} in Eq. \eqref{eq:ehat_1} enables to write an expression of the BEMFs estimates at equilibrium
\begin{equation}\label{eq:ehat_ab}
\left \{
\begin{array}{lcl}
\hat e_{\alpha} &=& -x\sin(\theta_e)+y\cos(\theta_e),\\
\hat e_\beta &=& x\cos(\theta_e)+y\sin(\theta_e),\\
x&=&p\omega(\phi_f-i_d^*\cdot \delta_L) -\delta_R\cdot i_q^*, \\
y &=& -i_d^*\cdot\delta_R+i_q^*\cdot p\omega \delta_L.
\end{array}\right.
\end{equation}
The steps leading to Eq. \eqref{eq:ehat_ab} are detailed in Appendix. 
From trigonometric identities, one can write
\begin{equation}\label{eq:eab_hat}
\hat e_{\alpha\beta}=\sqrt{x^2+y^2} \begin{bmatrix}
-\sin{(\theta_e+\varphi)}\\\cos{(\theta_e+\varphi)}
\end{bmatrix}                                                                                                                                                                                                                                                                                      
\end{equation}
with the angle $\varphi$ given at steady state by
\begin{equation}\label{eq:varphi}
\varphi^*(i_{dq},\delta_R,\delta_L,\omega) = \left\{
    \begin{array}{ll}
        -\arctan\left(\frac{y}{x}\right) & \mbox{if } x>0 \\
        -\arctan\left(\frac{y}{x}\right)+\pi & \mbox{otherwise}
    \end{array}
\right.
\end{equation}
Reminding Eq. \eqref{eq:eab}, the angle $\varphi$ appears to be the phase lag between $\hat e_{\alpha\beta}$ and $e_{\alpha\beta}$.

Eqs. \eqref{eq:ehat_ab} to \eqref{eq:varphi} give an analytic description of the \ac{BEMF} observation error in function of the uncertainties $\delta_R$, $\delta_L$ on the parameters $R$ and $L$. According to the model, the estimated BEMFs are delayed by a phase angle $\varphi$ and also have a biased amplitude. As mentioned previously, the amplitude bias will not alter the Park transformation. As for the phase lag, it will bias the Park transformation, leading to a modified rotating frame $\hat d-\hat q$. 
From Eqs. \eqref{eq:ehat_ab}, the phase lag $\varphi$ between the original $d-q$ frame and the observer $\hat d-\hat q$ frame is not constant since it depends on the machine currents $i_{dq}$ and the rotational velocity $\omega$. This phenomenon is analog to a misaligned encoder \cite{delpoux_joint_2012}, which misalignment would be dynamically depending on the machine operating points and the parameter uncertainties. 

Since the currents controller will operate within the biased $\hat d- \hat q$ frame, an error will be introduced in the control of the currents $i_{dq}$. 
Hence, extra analysis is needed to determine on which true $i_{dq}$ currents the system will stabilize. 

\subsection{A model of the currents equilibrium}

In this section, the goal is to develop a simple analytical model of the equilibrium adopted by the true $i_{dq}$ currents, given that their projections $ \hat i_{dq}$ are controlled within the biased $\hat d- \hat q$ frame. 
Let us write the projection of $\hat i_{dq}$ in the $d-q$ frame

\begin{equation}\label{eq:idq_problem}\left\{
\begin{array}{lcl}
i_d &=& \hat i_d\cos(\varphi) - \hat i_q\sin(\varphi),\\
i_q &=& \hat i_d\sin(\varphi) + \hat i_q\cos(\varphi).
\end{array}\right.
\end{equation}
From Eq. \eqref{eq:idq_problem}, it is possible to show that
\begin{equation}\label{eq:norm_i}
i_d^2+i_q^2=\hat i_{d}^2+\hat i_{d}^2,
\end{equation}
which means that the amplitude of the current vector is conserved from one frame to the other. From Eqs. \eqref{eq:varphi} and \eqref{eq:ehat_ab}, it could be noticed that the resolution of Eq. \eqref{eq:idq_problem} with respect to $i_{dq}$ is not trivial.  Hence, the following reasonable assumptions will be made.

\begin{assumption}{3}{}\label{a:omega}
The rotational velocity $\omega$ is positive.
\end{assumption}
The case where $\omega$ is very small or null is not considered since this would result in the system being non-observable, removing the purpose of the analysis.
\begin{assumption}{4}{}\label{a:torque}
High torque values take place for high velocities. 
\end{assumption}
This assumption holds since through \ac{OTC}, the currents follow the trajectory described by Eq. \eqref{eq:idqref}.
\begin{assumption}{5}{}\label{a:magn}
The parameter uncertainties $\delta_L$ and $\delta_R$ are of the same order of magnitude that the parameters $L$ and $R$. 
\end{assumption}
Given eq. \eqref{eq:idqref} and Assumption \ref{a:magn}, one can write that $\phi_f\gg i_d\cdot\delta_L$, unless uncommon PMSG design where the PM flux would be very weak. This condition is verified for the PMSG parameters provided in Table \ref{tab:parameters} since $\phi_f\simeq 10^{-1}$ Wb and $L\simeq 10^{-3} $ H. Hence, let us write an approximation of $x$ as 
\begin{equation}\label{eq:a_prime}
x'= p\phi_f\omega -\delta_R\cdot i_q^* \simeq x.
\end{equation}

\noindent Assumptions \ref{a:torque} and \ref{a:magn} allow to write the inequality 
\begin{equation}
p\phi_f\omega>\delta_R\cdot i_q^*,
\end{equation}
meaning that the term $a'$ of Eq. \eqref{eq:a_prime} is positive. Given the same assumptions, the term $b$ should stay very small compared to $x'$ leading to
\begin{equation}
x'\gg y.
\end{equation}  
From this, one can assume that the phase lag $\varphi$ is small. Consequently, the trigonometric functions could be reasonably approximated by the first term of their Taylor expansion. Hence, the phase lag $\varphi$ could be given by
\begin{equation}
\varphi  = -\arctan{\left( \dfrac{y}{x}\right) } \simeq  -\dfrac{y}{x'},
\end{equation}
and its projections in the trigonometric frame can be given by
\begin{equation}\label{eq:taylor_phi}
\begin{array}{lcl}
\sin{(\varphi)} & \simeq & -\frac{y}{x'}, \\
\cos{(\varphi)} & \simeq & 1.
\end{array}
\end{equation}
By injecting Eq. \eqref{eq:taylor_phi} in \eqref{eq:idq_problem}, the problem can be written at equilibrium such as 
\begin{equation}\label{eq:idq_simplified}
\begin{array}{lcl}
i_d^* &\simeq& \hat i_q^*\cdot \frac{y}{x'},\\
i_q^* &\simeq& \hat i_q^* .
\end{array}
\end{equation}
At equilibrium, the resolution of Eq. \eqref{eq:idq_simplified} leads to a simple expression of $i_d$ 
\begin{equation} \label{eq:id_model}
i_d^* \simeq \dfrac{\delta_L \,{i_q^\# }^2 }{\phi_f }.
\end{equation}
From Eq. \eqref{eq:idq_simplified} one can write that $\text{sign}(i_q)=\text{sign}(i_q^\#)$. Bearing in mind that the norm of the current is not changing from $d-q$ to $\hat d- \hat q$ frame, a more accurate approximation of $i_q$ can be found by injecting the approximation of $i_d$ found above in Eq. \eqref{eq:norm_i}, which leads to
\begin{equation} \label{eq:iq_model}
i_q^* \simeq  \text{sign}(i_q^\#)\cdot \sqrt{\frac{\phi_f \,\sqrt{4\,{\delta_L }^2 \,{ i_q^\# }^2 +{\phi_f }^2 }-{\phi_f }^2 }{2\,{\delta_L }^2 }}
\end{equation}
Hence, Eqs. \eqref{eq:id_model} and \eqref{eq:iq_model} give an analytical model of the true $i_{dq}$ currents equilibrium in function of the uncertainty on the parameters and the $i_q^\#$ current reference. It is interesting to note that according to the model, the parametric uncertainty $\delta_R$ does not introduce any error on $i_{dq}$. Hence, this model can be used to quantify the deviation of the system from the desired operating points in function of the parametric uncertainty $\delta_L$. Of course, using Eq. \eqref{eq:iq_model} with $\delta_L=0$ is of no interest.

\section{Results and discussion} \label{sec:results}

\begin{figure}[h]
\begin{center}
\includegraphics[width=0.4\textwidth]{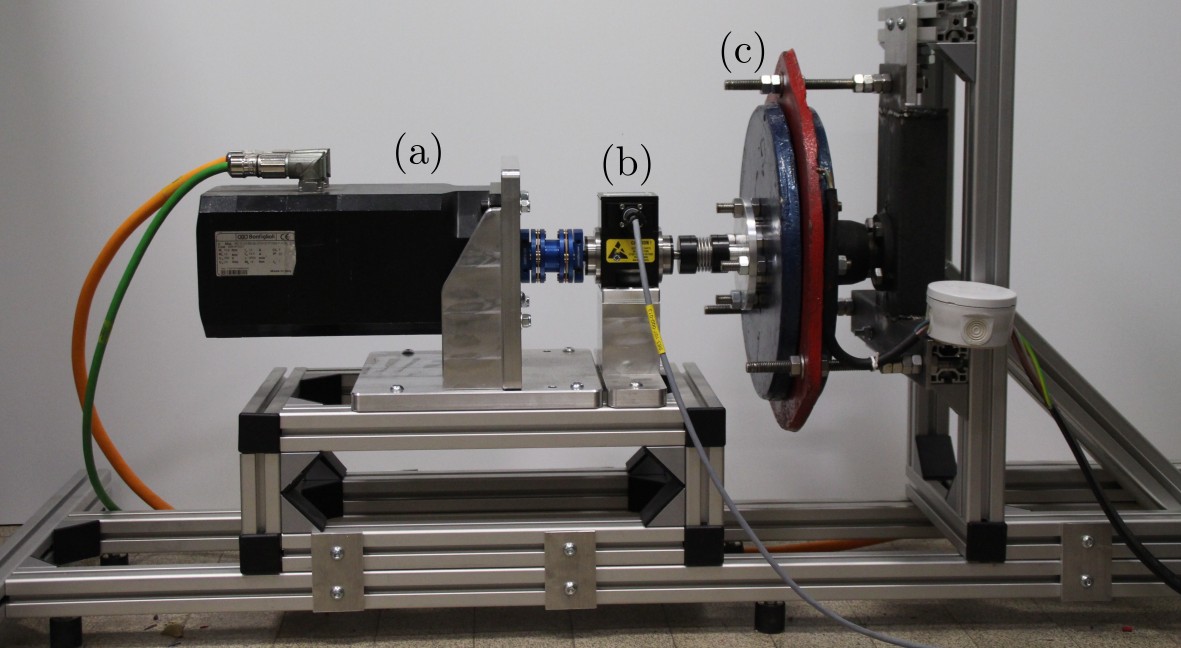}
\caption{Wind turbine emulator test bench. (a) PMSM (b) Torque-meter (c) axial flux PMSG}
\label{fig:banc}
\end{center}
\end{figure}

\begin{figure}[b]
\begin{center}
\includegraphics[width=0.45\textwidth]{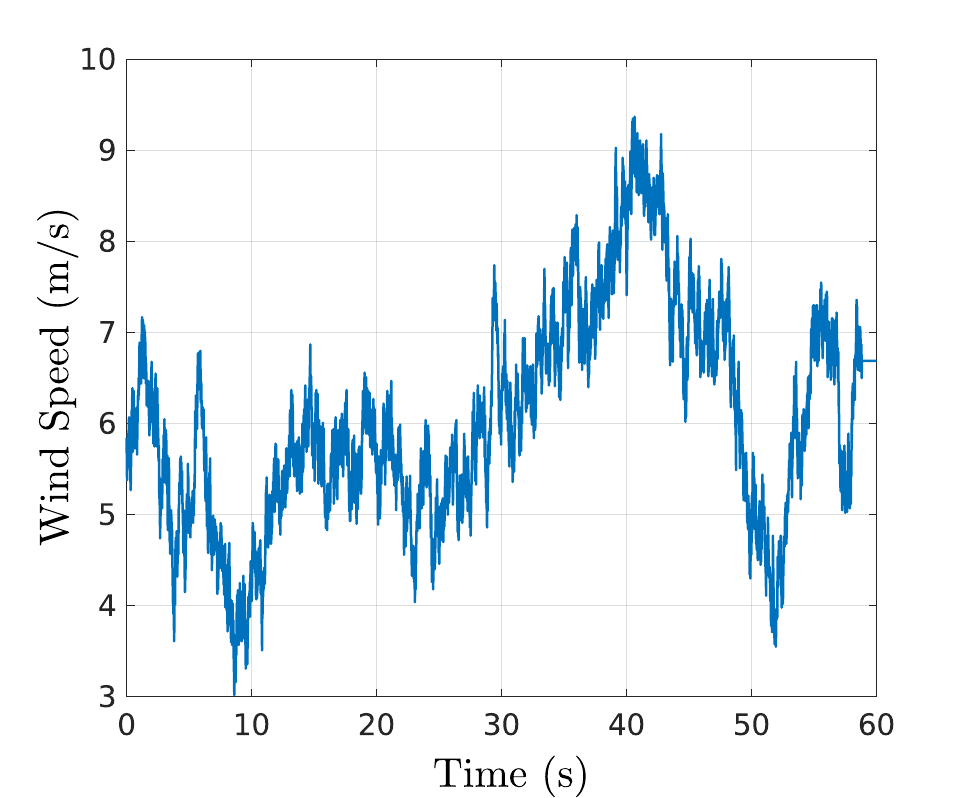}
\caption{The turbulent wind velocity time series}
\label{fig:wind}
\end{center}
\end{figure}

Experiments were conducted on the wind turbine emulator \cite{prevost_emulator_2023} shown on Fig. \ref{fig:banc}. The latter is made of a \ac{PMSG} and a \ac{PMSM} connected by their shaft through a torque meter. The \ac{PMSM} is controlled 
so that the torque of the turbine blades is emulated.
A first experiment was conducted under turbulent wind conditions, reported on Fig. \ref{fig:wind}, generated with TurbSim software using a Kaimal spectrum and an average wind velocity of 6 m/s. Tests was conducted with the same wind data first with an encoder and with the observer. For each of these tests, different values of the parameters {$L_o$, $R_o$} were provided. 
A second experiment was conducted with stationary wind conditions. This allows to measure the impact of parameter uncertainties on the power output over the whole wind speed range $[0-10]$ m/s of the "zone 2" operating zone. 
All the parameters used in the experimental setup are summarized in Table \ref{tab:parameters}. The aerodynamic coefficient $C_p(\lambda)$ is issue from
 \cite{national_and_technical_university_of_athens_aerodynamic_2023}. 

\subsection{Tracking performance}

\subsubsection{Currents}
\begin{figure}[t]
\begin{center}
\includegraphics[width=0.45\textwidth]{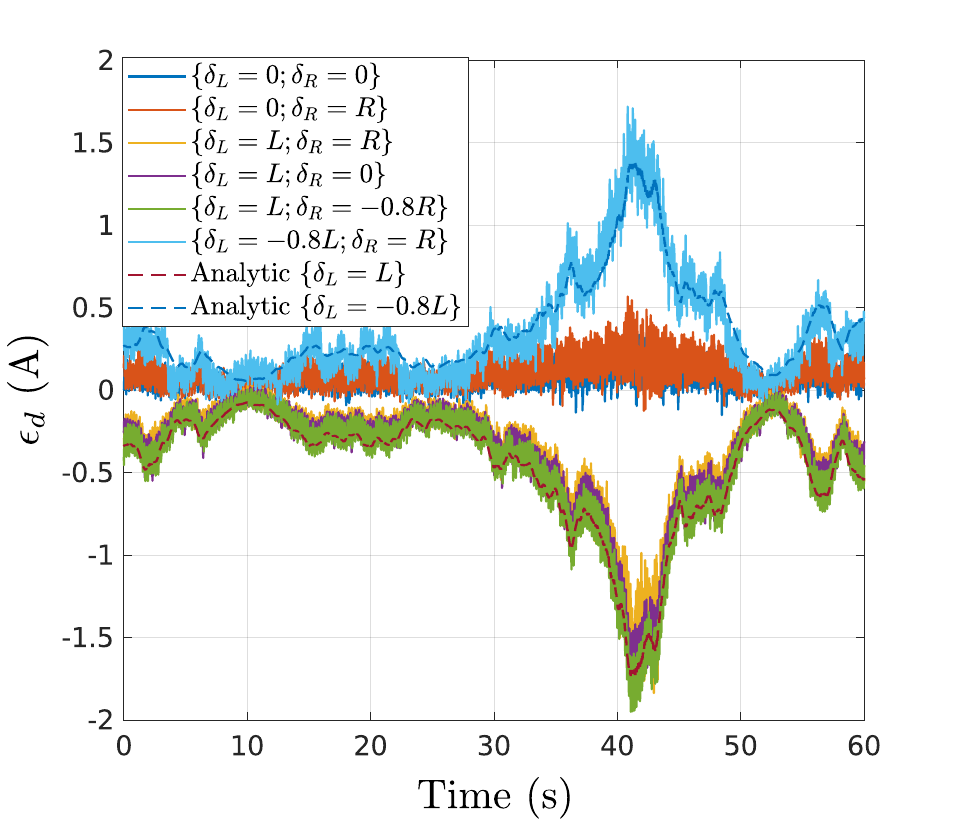}
\caption{Tracking error of the current $i_d$ for different scenario of parameters uncertainty. Experimental data are plotted in continuous lines and results from the analytical model in dashed.}
\label{fig:id}
\end{center}
\end{figure}

\begin{figure}[h]
\begin{center}
\includegraphics[width=0.45\textwidth]{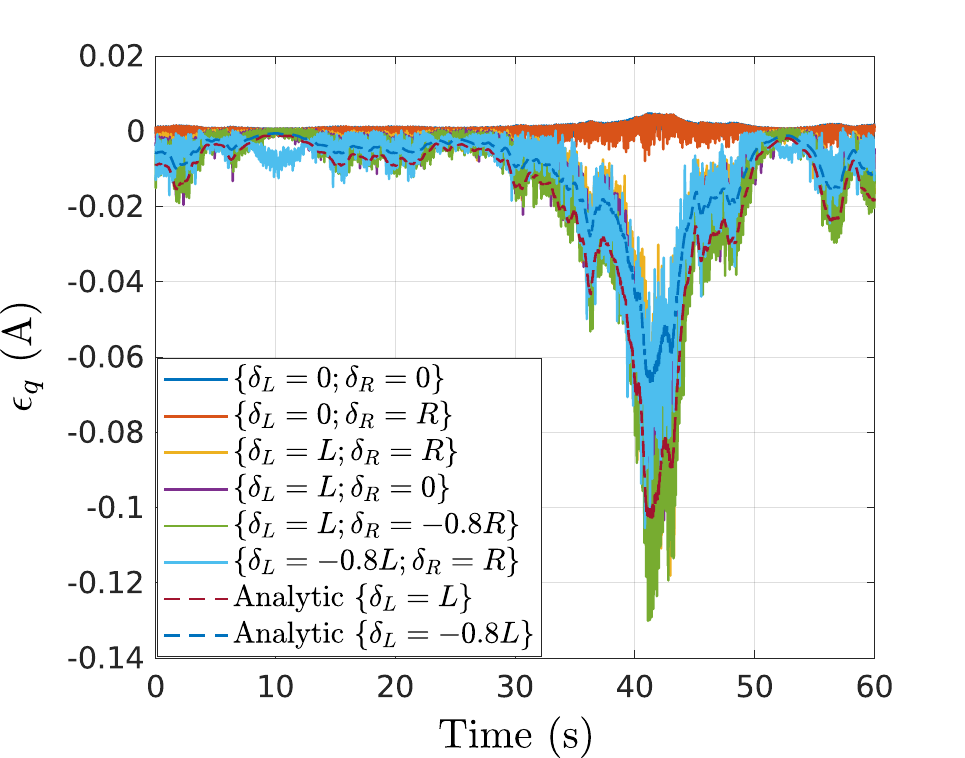}
\caption{Tracking error of the current $i_q$ for different scenario of parameter uncertainties. Experimental data are plotted in continuous lines and results from the analytical model in dashed.}
\label{fig:iq}
\end{center}
\end{figure}

The currents tracking errors $\epsilon_{dq}=e_{dq}=i_{dq}^\#-i_{dq}$ are plotted on Figs. \ref{fig:id} and \ref{fig:iq} respectively for each parameter uncertainties scenario. It can be seen that the tracking error is close to zero when the correct inductance value is provided to the observer (i.e.\ $\delta_L=0$). However, when $\delta_L\neq 0$, it can be observed that a tracking error is introduced. 
The analytical tracking errors from Eqs. \eqref{eq:id_model} and \eqref{eq:iq_model} are plotted in dashed lines for the scenario $\delta_L=-0.8L$ and $\delta_L=L$. On Fig. \ref{fig:id}, one can see that the tracking error amplitude $\epsilon_d$ rises when the wind velocity increases. Indeed, as wind velocity rises, the aerodynamic torque increases, leading to a surge of the rotor speed which, through \ac{OTC} (Eq. \eqref{eq:idqref}) raises the current setpoint $i_q^\#$ amplitude. According to Eq. \eqref{eq:id_model}, this causes an increase of the $i_d$ current amplitude, despite the fact that $i_d^\#=0$. It is interesting to note that a non-desired flux weakening occurs in the cases where $\delta_L<0$, which is coherent with the analytical model. It can be seen that the evolution of $\epsilon_d$ computed with the model fits well with the experimental data in average value. Despite up to 2A of negative current on the direct current $i_d$, the overall energy harvest, shown Fig.~\ref{fig:W}, was not impacted significantly. Indeed, the currents' norm $||i_{dq}||$ as mentioned in \eqref{eq:norm_i}, remains unchanged. It leads to no extra joule loss but a slight torque production reduction, see Eq. \eqref{eq:iq_model}, which does not affect the aerodynamic operating point of the generator, see Fig.~\ref{fig:lambda}. 
The tracking error $\epsilon_q$ represented on Fig. \ref{fig:iq} is also close to zero when the uncertainty on the inductance is zero, which means that the reference torque computed by OTC is correctly tracked. In the cases where $\delta_L\neq 0$, negative tracking errors were recorded which amplitudes also increase with wind speed for the reasons explained above. This is also well-fitted in average by the curves computed from the analytical model described by Eq.  \eqref{eq:iq_model}. This negative tracking error on $i_q$ results in a part of the torque being lost in reactive power. 

\subsubsection{Rotational velocity} 
\begin{figure}[h]
\begin{center}
\includegraphics[width=0.45\textwidth]{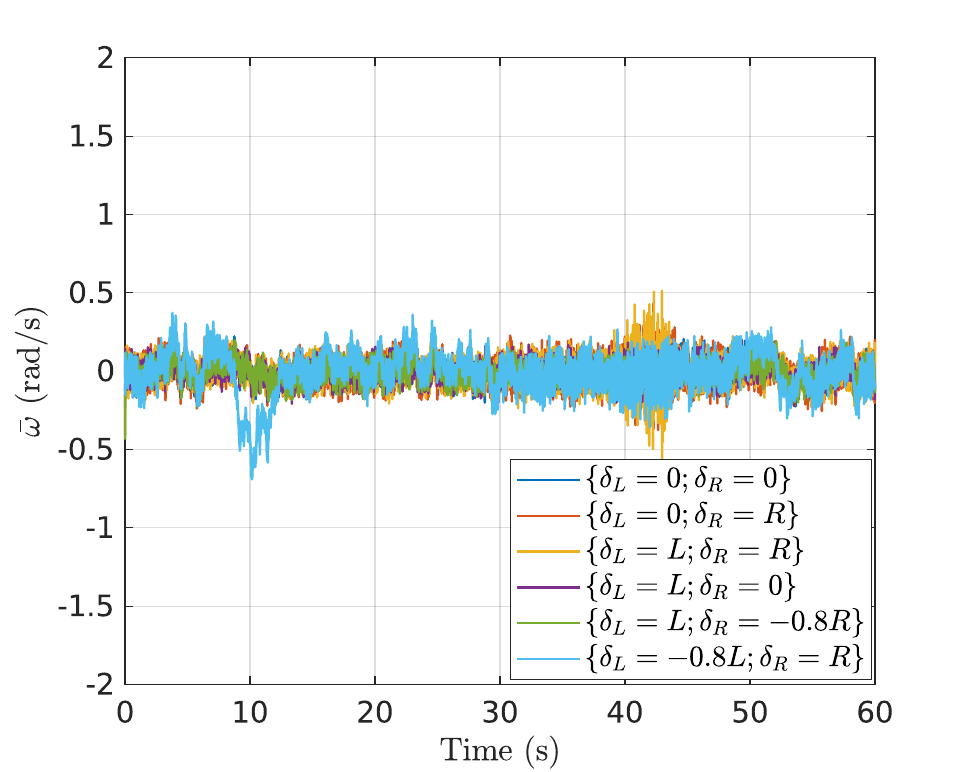}
\caption{Speed observation error for several scenario of parameters uncertainty in turbulent wind conditions.}
\label{fig:om}
\end{center}
\end{figure}

The speed tracking error $\bar \omega=\hat \omega - \omega$ is represented on Fig. \ref{fig:om}. It stays close to zero for all uncertainties scenarios, which is crucial since the rotational velocity is used by OTC to compute the currents references.

\subsubsection{Tip Speed Ratio} 
\begin{figure}[h]
\begin{center}
\includegraphics[width=0.45\textwidth]{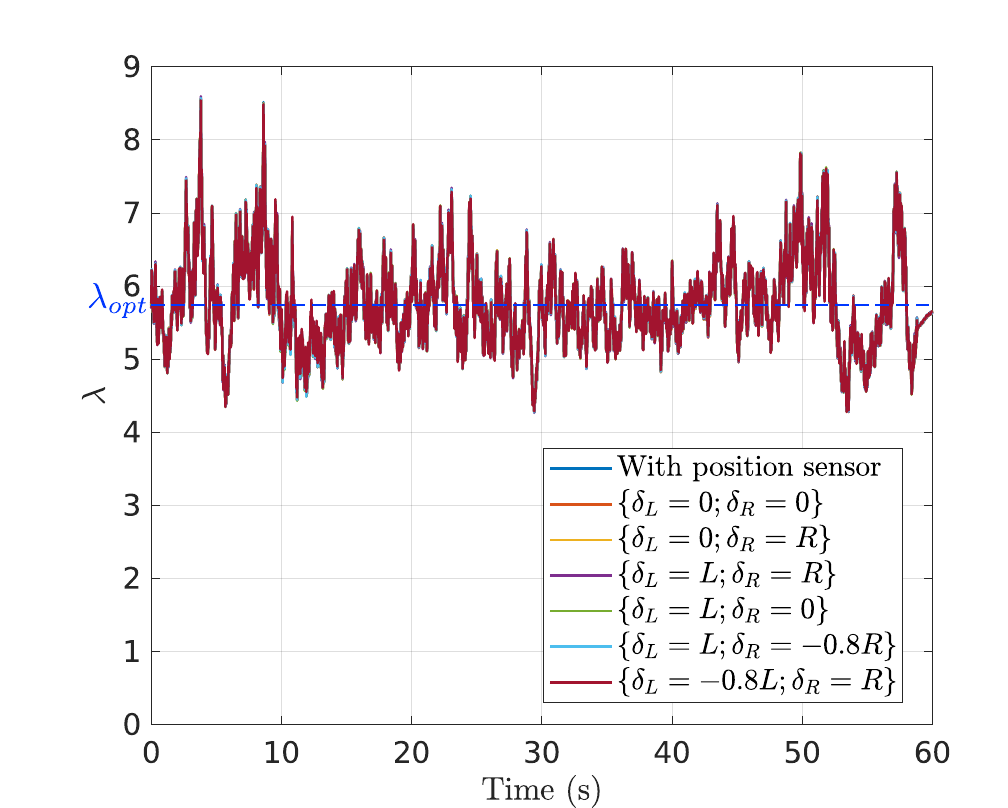}
\caption{Effect of the uncertainties on the TSR for turbulent wind.}
\label{fig:lambda}
\end{center}
\end{figure}

The TSR is plotted on Fig. \ref{fig:lambda}. 
It can be seen that parameter uncertainties have no impact on it since it stays close to the optimal TSR $\lambda_{opt}=5.75$. It shows that OTC is not affected by the parameter uncertainties. Indeed, since the tracking errors on the currents and the rotational velocity are small, the impact on the generator torque is not significant. This results in identical aerodynamic behaviors.

\subsection{Energy harvesting}
\begin{figure}[h]
\begin{center}
\includegraphics[width=0.45\textwidth]{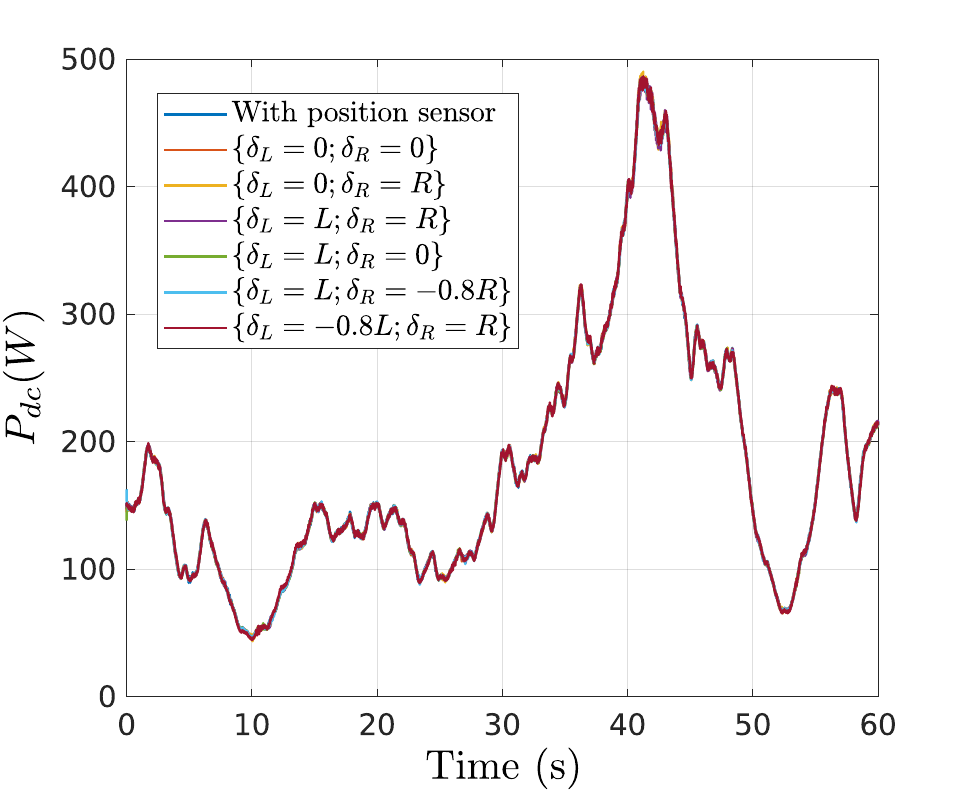}
\caption{Power at the rectifier output for several system configurations and scenario of parameters uncertainty.}
\label{fig:P_dynamic}
\end{center}
\end{figure}

\begin{figure}[h]
\begin{center}
\includegraphics[width=0.45\textwidth]{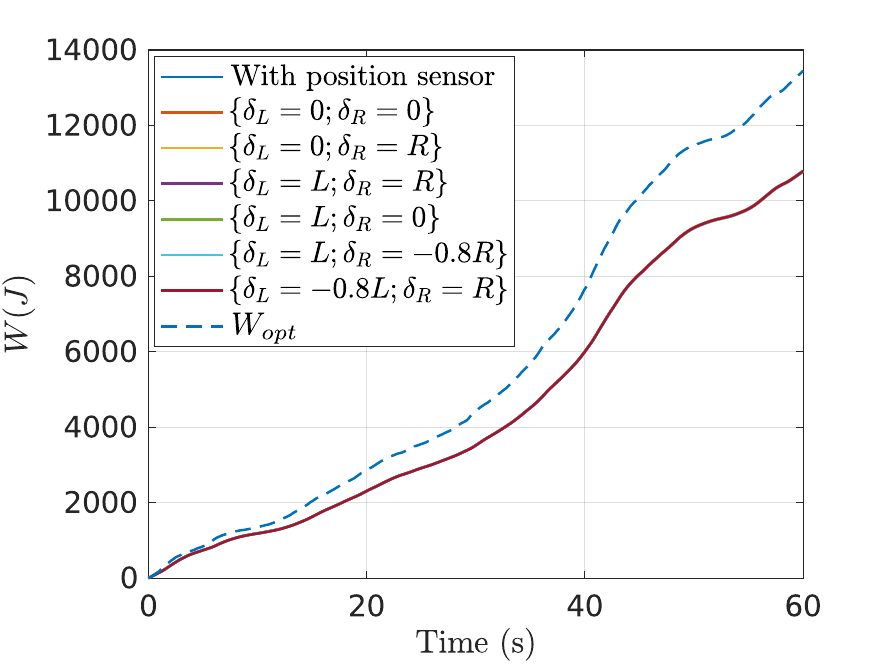}
\caption{WECS energy yield.
}
\label{fig:W}
\end{center}
\end{figure}

It can be seen on Fig. \ref{fig:P_dynamic} that for dynamic conditions, the uncertainty on parameters does not affect the output power. This result is not surprising since the tracking error on $i_q$ was shown to remain small in spite of high uncertainty on the parameters, which leads to close aerodynamic operating points as seen with the TSR on Fig. \ref{fig:lambda}. It should be reminded that only currents references are used to control the generator and that the norm of the currents are the same in both $d-q$ and $\hat d-\hat q$ frames. Hence, an angle observation error shall not introduce additional joules losses in comparison to the encoder scenario. 
Let us define the optimal energy harvest as
\begin{equation}
W_{opt}=\int_0^T \dfrac{1}{2}\rho_{air} A V_w(t)^3 C_{p(max)} dt
\end{equation} with $T$ the duration of the wind velocity time series. This energy corresponds to an ideal scenario where the aerodynamic efficiency would stay always optimal (i.e.\ $C_p=C_{p(max)}$) while the generator-rectifier efficiency would be unity. The energetic efficiency $\eta_E$ over the period $T$ is given by
\begin{equation}
\eta_E=\dfrac{\int_0^T P_{dc}(t)dt}{W_{opt}}=\dfrac{W}{W_{opt}}.
\end{equation}
The energy harvest $W$ for each system configuration and parameter uncertainties scenario together with the optimal energy harvest $W_{opt}$ are plotted on Fig. \ref{fig:W}. A summary of the energy yield is shown on Table \ref{tab:results_summary}. The control strategy allowed to harvest $80\%$ of available energy $W_{opt}$ despite parameter uncertainties.
\begin{table}[h!]
\begin{center}
\begin{tabular}{@{}llllll@{}}
\toprule
Configuration & $\delta_L$    & $\delta_R$    & AEP (kWh)     & $\eta_E$ \\ \midrule
Encoder       & -     & -     & 1183     & 0.80     \\
Sensorless      & 0     & 0     & 1167  & 0.80     \\
Sensorless      & 0     & R     & 1167  & 0.80     \\
Sensorless      & L     & R     & 1164  & 0.80     \\
Sensorless      & L     & 0     & 1166  & 0.80     \\
Sensorless      & L     & -0.8R & 1163  & 0.80     \\
Sensorless      & -0.8L & R     & 1162  & 0.80     \\ \bottomrule
\end{tabular}
\caption{Performance's Summary of the control strategy.}
\label{tab:results_summary}
\end{center}
\end{table}

\subsection{Annual Energy Production}
\begin{figure}[h]
\begin{center}
\includegraphics[width=0.45\textwidth]{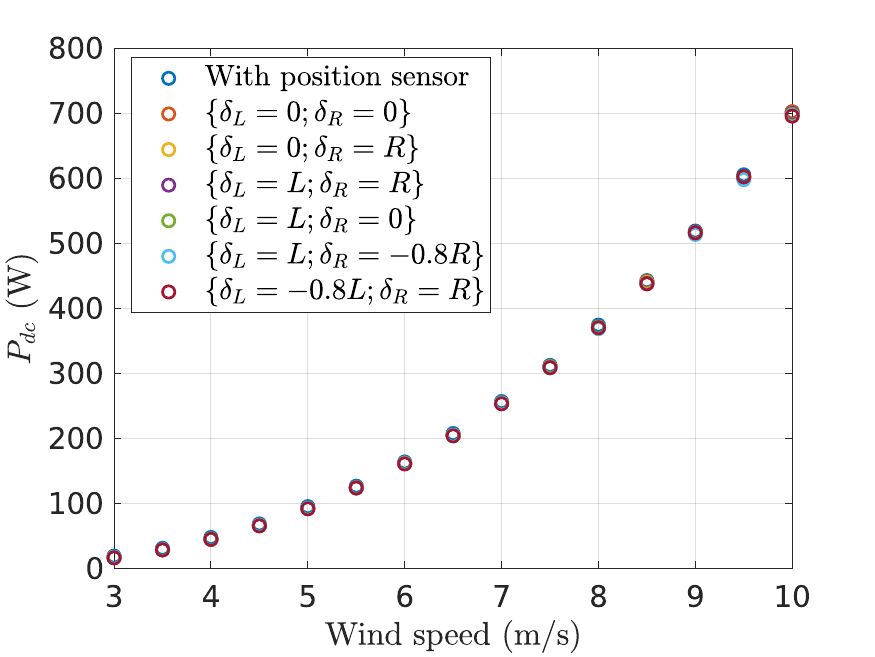}
\caption{Comparison of power curves of the WECS under for several scenario of parameter uncertainties.}
\label{fig:P_stat}
\end{center}
\end{figure}

The power curves measured during the experiments under stationary wind are represented on Fig. \ref{fig:P_stat}. The truncated \ac{AEP} was computed \cite{noauthor_en_2006} for a Rayleigh distribution with an average wind velocity of 5 m/s. Amongst the different scenario of parameter uncertainties, the AEP is at least $98\%$ of the AEP obtained with an encoder.

\section{Conclusion}

In this work, a mechanical sensorless control strategy for small wind turbines with a surface mounted PMSG was designed and analyzed regarding parameter uncertainties. Under a constant or slowly varying blades torque, we proved that the system is \ac{GAS}. Additionally, a condition was provided on the SMO gain to ensure convergence in spite of the uncertainties. The modelling of the steady state tracking error of the machine currents 
allowed to illustrate the robustness of the proposed approach. Experiments where conducted on a wind turbine emulator to investigate how the tracking errors would impact the behavior of the WECS.
The analytical model gave a good approximation of the experimental average tracking errors when compared to experimental data. 
In both stationary and turbulent wind conditions, the sensorless strategy performance was similar to the configuration with an encoder, in spite of the uncertainties on the parameters. Hence, sensorless control of PMSGs does not necessarily degrade the power performance of WECSs under active rectification. 
Overall, the algorithm could be implemented on a standard synchronous rectifier controller with a 'generic' set of electrical parameters. In future works, it may be interesting to extend the control law to the zone III and investigate how parameter uncertainties affect its behavior in the stall region.

\color{black}


\bibliographystyle{Bibliography/IEEEtranTIE}
\bibliography{Bibliography/IEEEabrv,Bibliography/references}\ 

\section*{Appendix}

\begin{table}[ht!]
\begin{center}
\begin{tabular}{lll|lll}
\toprule
\textbf{Parameter}           & \textbf{Value} & \textbf{Unit} & \textbf{Parameter}           & \textbf{Value} & \textbf{Unit}   \\ \midrule
$P_{nom}$ & 700 & W\\
$V_{dc}$                        & 50             & V  &             
$I_{max}$ & 20&A\\
$p$  & 8 & - &
$R_s$                       & 0.42           & $\Omega$            \\
$L_s$                           & 1            & mH              &
$\phi_f$                         & 0.11           & Wb              \\
$J$							& 0.66			& kg.m$^2$ &
$b$                         & 0.008          & N.m.s/rad      \\
$C_{p(max)}$                        & 0.33           &                 &
$\rho_{air}$ & 1.204          & $kg/m^3$           \\
$\lambda_{opt}$				& 5.75	& &
$R_r$                           & 1.2            & m        \\ 
$K_{opt}$ & $0.0088$ & &
$l_1$ & $30$& \\
$l_2$ & $100$& &
$l_3$ & $10$& \\
 \bottomrule
\end{tabular}
\end{center}
\caption{Summary of the experimental setup system parameters.}
\label{tab:parameters}
\end{table}

\subsection*{Proof of Theorem \ref{t:syst_stability}}
The system described by \eqref{eq:cl_system} which state vector is $x=[e_d,e_q,e_{i_d},e_{i_q},e_{\omega}]^T\in\mathbb{R}^5$ has a unique equilibrium 
\begin{equation}
    x^*=\left[0,0,0,0,0\right]^T.
\end{equation}
A Lyapunov function candidate is proposed such as 
\[
\left\lbrace
\begin{array}{lll}
V &= & \beta(V_d+V_q) +V_{\omega} \\
&&\\
V_d &=&\frac{1}{2}Le_d^2+\frac{1}{2}k_ie_{id}^2\\
&&\\
V_q &=&\frac{1}{2}Le_q^2+\frac{1}{2}k_ie_{iq}^2\\
&&\\
V_{\omega} & =& \frac{1}{2}e_{\omega}^2\\
\end{array}
\right.
\]
with $\beta>0$ and $k_i>0$. The derivative in the $d$ axis is given by
\begin{equation}
    \begin{array}{lll}\label{eq:vddot}
\dot{V}_d &=&Le_d\dot{e}_d+k_ie_{id}\dot{e}_{id}\\
    &=& -(k_p+R){e^2_d}+pL(e_{\omega}+\omega^*)e_qe_d+pLe_{\omega} {e_d}{i^{\#}_{q}}\\
\end{array}
\end{equation}
and in the $q$ axis by
\begin{equation}\label{eq:vqdot}
    \begin{array}{lll}
\dot{V}_q &=&Le_q\dot{e}_q+k_ie_{iq}\dot{e}_{iq}\\
     &=& -(k_p+R){e^2_q}-pL(e_{\omega}+\omega^*)e_de_q\\&&-p(\phi_f+L{i^{\#}_{d}})e_{\omega}{e_q}.\\
\end{array}
\end{equation}
Hence, 
\begin{equation}
    \begin{array}{lll}
\dot{V}_d+\dot{V}_q   & = & -(k_p+R){e^2_d}-(k_p+R){e^2_q}+pLe_{\omega} {e_d}{i^{\#}_{q}}\\&& -p(\phi_f+L{i^{\#}_{d}})e_{\omega}{e_q}.\\
\end{array}
\end{equation}
Moreover, the derivative for $V_\omega$ can be given by
\begin{equation}\label{eq:vomegadot}
    \begin{array}{lll}
\dot{V}_{\omega}  &=&e_{\omega}\dot{e}_{\omega}\\
                  &=&\frac{\frac{3}{2}p\phi_f}{J} {e_{\omega}}{e_q}+\frac{\frac{3}{2}p\phi_f}{J}{e_{\omega}}{i^{\#}_{q}}-\frac{b}{J}{e^2_{\omega}}-\frac{b}{J}{e_{\omega}}{\omega^*}.\\
\end{array}
\end{equation}
Since $\omega^*=\frac{3}{2b}p\phi_fi^{\#}_q$, eq. \eqref{eq:vomegadot} can be simplified such as 
\begin{equation}\label{eq:vomegadot_simple}
    \begin{array}{lll}
\dot{V}_{\omega} &=&\frac{\frac{3}{2}p\phi_f}{J} {e_{\omega}}{e_q}-\frac{b}{J}{e^2_{\omega}}.\\
\end{array}
\end{equation}
From eqs. \eqref{eq:vddot}, \eqref{eq:vqdot}, \eqref{eq:vomegadot_simple}, the Lyapunov function derivative can be written as
\begin{equation}
    \begin{array}{lll}
\dot{V} & = & -(k_p+R)\beta{e^2_d}-(k_p+R)\beta{e^2_q}-\frac{b}{J}{e^2_{\omega}}+pL\beta{e_{\omega}} {e_d}{i^{\#}_{q}}\\&&-p\beta(\phi_f+L{i^{\#}_{d}})e_{\omega}{e_q} +\frac{\frac{3}{2}p\phi_f}{J} {e_{\omega}}{e_q}\\
\end{array}
\end{equation}
which can be re-written as 
\begin{equation}
    \dot{V} =
\begin{bmatrix}
e_d & e_q & e_{\omega} \\
\end{bmatrix}
\underbrace{
\begin{bmatrix}
 -a\beta & 0 & \beta d \\
  0& -a\beta &  e-\beta c\\
  \beta d &  e-\beta c & -\frac{b}{J}\\
\end{bmatrix}}_{M}
\begin{bmatrix}
e_d \\
e_q \\
e_{\omega}\\
\end{bmatrix}
\end{equation}
with $a=k_p+R$, $c=p({\phi_f}+Li^{\#}_{d})/2$, $d=\frac{1}{2} pLi_q^{\#}$ and $e=\frac{\frac{3}{2}p\phi_f}{2J}$.
The determinant is given by
\begin{equation}
\begin{array}{lll}
   \det M & = & -\frac{\gamma+a\beta}{J}\left(J\gamma^2+(b+Ja\beta)\gamma\right.\\&&\left.{-J(e^2+\beta^2c^2+\beta^2d^2)+2Jce\beta+ab \beta} \right).
\end{array}
\end{equation}
implying three eigenvalues $\gamma_1, \gamma_2, \gamma_3$. The first eigenvalue is written as
\begin{equation}
    \gamma_1=-a\beta.
\end{equation} 
Since $\beta>0$, the following condition must be met to ensure that the first eigenvalue is negative
\begin{equation}
    k_p>-R.
\end{equation}
Given this condition, the coefficients factors of $\gamma$ and $\gamma^2$ are positive, since $J$ and $b$ are positive parameters. Thus, $\gamma_2, \gamma_3$ are negative if
\begin{equation}
    P(\beta)=-J(e^2+\beta^2c^2+\beta^2d^2)+2Jce\beta+ab \beta >0.
\end{equation}
Since $P(\beta)$ is a downwards parabola, it can be positive if $\max P(\beta)>0$. One can write 
\begin{equation}
    \frac{d P}{d\beta} = -2J(c^2+d^2)\beta+2Jce+ab
\end{equation}
hence, $\beta_{max} = \frac{2Jce+ab}{2J(c^2+d^2)}$ and 
\begin{equation}
    P(\beta_{max}) = \dfrac{1}{4J(c^2+d^2)}Q(a)
\end{equation}
with $Q(a)=-4\,J^2 \,d^2 \,e^2 +4\,c\,J\,a\,b\,e+a^2 \,b^2$ which is an upwards parabola with two roots $\{a_1, a_2\}\in \Re^2 $ with $a_2>0>a_1$ and 
\begin{equation}\label{eq:a2}
    a_2= \frac{2\,{\left(J\,e\,\sqrt{c^2 +d^2 }-J\,c\,e\right)}}{b}.
\end{equation}
Inserting the expressions of $c,d,e$ in eq. \eqref{eq:a2} leads to 
\begin{equation}
    a_2=\dfrac{3}{4b}p\phi_f\sqrt{p^2(\phi_f+Li_d^\#)^2+(L{i_q^\#}p)^2}-\dfrac{3p^2\phi_f(\phi_f+Li_d^\#)}{4b}.
\end{equation}
Hence, one can write that
\begin{equation}
    a>a_2\Rightarrow Q(a)>0 \Rightarrow P(\beta_{max})>0\Rightarrow\{\gamma_2,\gamma_3\}<0
\end{equation}
In a nutshell, the following conditions
\begin{equation}
    \left\lbrace
\begin{array}{lll}
k_p & >&-R \\
k_p &>& a_2-R\\
\end{array}
\right.
\end{equation}
shall be met to guarantee that $\dot V <0$. Since $R$ is a positive physical parameter and $a_2>0$ the dominant condition is
\begin{equation}\label{eq:stab_condition}
k_p > a_2-R.
\end{equation} If the condition \eqref{eq:stab_condition} is satisfied, \( \dot{V} \leq 0 \quad \forall x \in \Re^5 \) which proves that the equilibrium \( x^*=0 \) is globally stable. Let \( S \) be the set of solutions where \( \dot{V} = 0 \), i.e., \( S = \{ x \in \Re^5 \mid e_d = e_q = e_{\omega} = 0 \} \). The unique invariant solution of \( S \) is \( x^*=0 \) thus, according to LaSalle's theorem, it can be concluded that the convergence is asymptotic to the equilibrium point. Finally, since $V>0$ for all $x\neq 0$ and $V(0)=0$ it results that the system is \ac{GAS}, which ends the proof.

\color{black}
\subsection*{Proof of Eq. \eqref{eq:l1_robust}} 
From Eq. \eqref{eq:l1_theo}, a slightly more restrictive condition is given by
\begin{equation}
\begin{array}{lcl}
l_1&>&\frac{L_o}{L}\cdot\text{max}(\mid e_\alpha\mid,\mid e_\beta\mid)+\text{max}(\mid \pi_\alpha\mid,\mid \pi_\beta\mid).\\
\end{array}
\end{equation}
Let us find a bound for the perturbation terms $\Pi_\alpha$ and $\Pi_\beta$
\begin{equation}
\begin{array}{lcl}
\mid \Pi_{\alpha} \mid &=&\mid \left( \frac{R\delta_L}{L}-\delta_R \right) i_{\alpha}-\frac{\delta_L}{L} v_{\alpha}\mid,\\
&\leq & \left( \frac{R\mid\delta_L\mid}{L}+\mid\delta_R\mid \right) I_{max} +\frac{\mid\delta_L\mid}{L}  V_{max},\\
&\leq &  \left( R_{max}\dfrac{\Delta_L}{L_{min}}-\Delta_R\right)I_{max}+\dfrac{\Delta_L}{L_{min}}V_{max}=\Pi_{max}
\end{array}.
\end{equation}

Applying the same reasoning for $\Pi_{\beta}$ gives rise to the same bound $\Pi_{max}$, which leads to
$\Pi_{max}\geq \text{max}(\mid \pi_\alpha\mid,\mid \pi_\beta\mid).$
Since
\begin{equation}
\dfrac{L_o}{L}e_{\alpha\beta}\leq \dfrac{L_{o}}{L_{min}} E_{max},
\end{equation}
the bound on $l_1$ naturally comes as 
\begin{equation}
l_1>\dfrac{L_{o}}{L_{min}} E_{max}+\Pi_{max},
\end{equation}
which proves the bound given by Eq. \eqref{eq:l1_robust}.

\textbf{Proof of Eq. \eqref{eq:ehat_ab}:} 
Using the inverse of Eq.~\eqref{eq:park} 
in Eq. \eqref{eq:ehat_1} gives
\begin{equation}\label{eq:eab_hat2}
\begin{array}{lcl}
\hat e_{\alpha} &=& -\frac{L_o}{L}p\phi_f\omega \sin(\theta_e)+\frac{R\delta_L-\delta_RL}{L} (i_d\cos(\theta_e)-i_q\sin(\theta_e))\\&&- \frac{\delta_L}{L} (v_d\cos(\theta_e)-v_q\sin(\theta_e)),\\
\hat e_\beta &=& \frac{L_o}{L}p\phi_f\omega \cos(\theta_e)+\frac{R\delta_L-\delta_RL}{L} (i_d\sin(\theta_e)+i_q\cos(\theta_e))\\&&-\frac{\delta_L}{L}(v_d\sin(\theta_e)+v_q\cos(\theta_e)).
\end{array}
\end{equation}
After manipulations, it leads to  
\begin{equation}
\begin{array}{lcl}
\hat e_{\alpha} &=& -x\sin(\theta_e)+y\cos(\theta_e),\\
\hat e_\beta &=& x\cos(\theta_e)+y\sin(\theta_e)
\end{array}
\end{equation}
with
\begin{equation}\label{eq:ab}
\begin{array}{lcl}
x&=&\frac{L_o}{L}p\phi_f\omega -\frac{\delta_L}{L} v_q+\frac{R\delta_L-\delta_RL}{L}i_q ,\\
y &=& -\frac{\delta_L}{L}v_d+\frac{R\delta_L-\delta_RL}{L}i_d .
\end{array}
\end{equation}

Injecting Eq. \eqref{eq:vdq_std} in Eq. \eqref{eq:ab} allows simplifying the terms $a$ and $b$ gives:
\begin{equation}
\begin{array}{lcl}
x&=&\frac{L+\delta_L}{L}p\phi_f\omega-\frac{\delta_L}{L}\left( R i_q^*+p\omega L  i_d^* +p\phi_f \omega \right)  \\&& + \frac{R\delta_L}{L}i_q^* - \frac{\delta_RL}{L}i_q^* ,\\
&=& p\omega(\phi_f-i_d^* \delta_L) -\delta_R i_q^*.\\
y&=& -\frac{\delta_L}{L}\left( R i_d^*-p\omega L  i_q^* \right) + \frac{R\delta_L}{L}i_d^* - \frac{\delta_RL}{L}i_d^*, \\
&=& -i_d^*\delta_R+i_q^* p\omega \delta_L
\end{array}
\end{equation}

which ends the proof.
	
\vspace{0.5cm}
\begin{IEEEbiography}[{\includegraphics[width=1in,height=1.25in,clip,keepaspectratio]{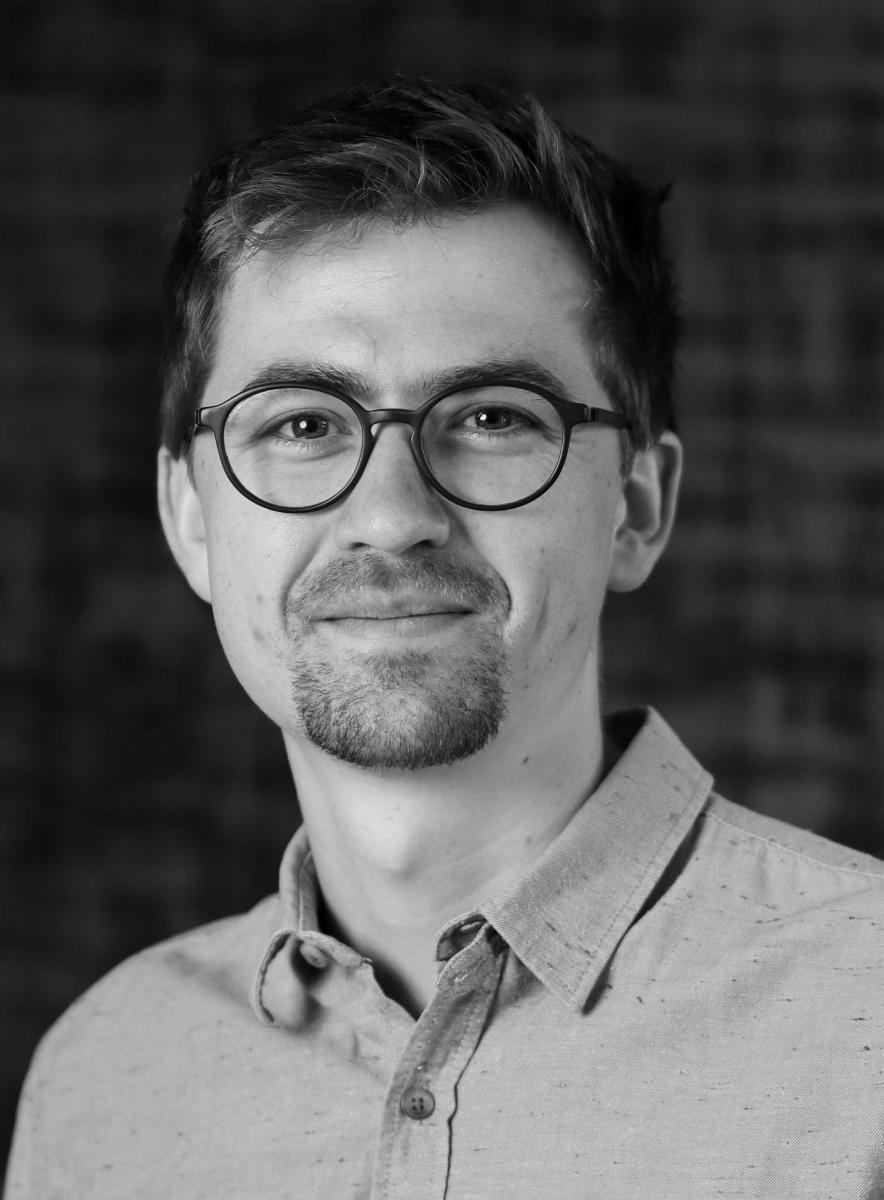}}]
{Adrien Pr\'evost} received the M.S. degree in electrical engineering from Institut National des Sciences Appliquées de Lyon (INSA Lyon), Villeurbanne, France, in 2020. He received the Ph.D. degree in control theory from INSA Lyon in 2024. He is currently a post-doc researcher at G2ELab, Institut National Polytechnique (INP) Grenoble, France. His research includes modeling, design, control and technico-environmental optimization of electrical machine-converter systems.
\end{IEEEbiography}

\vspace{-1cm}
\begin{IEEEbiography}[{\includegraphics[width=1in,height=1.25in,clip,keepaspectratio]{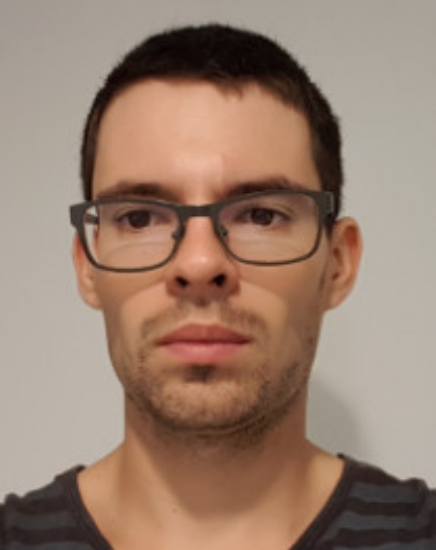}}]
{Vincent L\'echapp\'e} received the M.S. and Ph.D. degrees in control theory from the Ecole Centrale de Nantes, Nantes, France, in 2012 and 2015, respectively, both in control theory. He was a Lecturer with the School of Engineering and Physics, the University of the South Pacific, Suva, Fiji, in 2016. He joined Griffith University in Australia for a Postdoctoral stay, in 2017. In November 2017, he joined INSA Lyon, France, where he is currently an Associate Professor at Electrical Engineering Department. His research interests include control and observation of electrical machines and their environmental impacts.

\end{IEEEbiography}

\vspace{-1cm}
\begin{IEEEbiography}[{\includegraphics[width=1in,height=1.25in,clip,keepaspectratio]{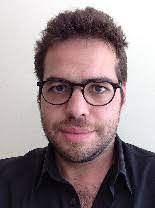}}]
{Romain Delpoux} received the M.S. degree in control system and mechatronics form the Chalmers University of Technology, Gothenburg, Sweden, in 2009 and the Ph.D. degree in control theory, data and signal processing from the \'Ecole Centrale de Lille, Villeneuve-d’Ascq, France, in 2012. In September 2014, he joined INSA Lyon, Villeurbanne, France, and the Ampere Laboratory, where he is currently an Associate Professor. His research interests include embedded control applications for power electronics systems.

\end{IEEEbiography}

\vspace{-1.9cm}
\begin{IEEEbiography}[{\includegraphics[width=1in,height=1.25in,clip,keepaspectratio]{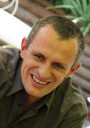}}]
{Xavier Brun} was born in France in 1973. He received the Ph.D. degree from the Institut National des Sciences Appliqu\'ees de Lyon (INSA
Lyon), Villeurbanne, France, in 1999. He became an Associate Professor with the Laboratoire d’Automatique Industrielle, INSA Lyon, in 2001. Since 2011, he has been a Professor at the Laboratoire Ampere, INSA Lyon, France. His current research interests include control for mechatronic and fluid power systems.
\end{IEEEbiography}

\end{document}

%% file: a_acronyms.tex
%
\begin{acronym}[RCP]
\acro{AEP}{Annual Energy Production}
\acro{AFPM}{Axial Flux Permanent Magnet}
\acro{BEMF}{Back Electromotive Force}
\acrodefplural{BEMF}{Back Electromotive Forces}
\acro{DPC}{Direct Power Control}
\acro{DR}{Diode Rectifier}
\acro{EGS}{Electrical Generation System}
\acro{ESC}{Extremum Seeking Control}
\acro{EROI}{Energy Return On Investment}
\acro{FOC}{Field Oriented Control}
\acro{GAS}{Globally Asymptotically Stable}
\acro{IPC}{Indirect Power Control}
\acro{LCOE}{Levelized Cost Of Energy}
\acro{LE}{Load Emulator}
\acro{LMSWT}{Locally Manufactured Small Wind Turbine}
\acro{MPPT}{Maximum Power Point Tracking}
\acro{OTC}{Optimal Torque Control}
\acro{PHIL}{Power Hardware In the Loop}	
\acro{PM}{Permanent Magnet}
\acrodefplural{PM}{Permanent Magnets}	
\acro{PMSG}{Permanent Magnet Synchronous Generator}
\acro{PMSM}{Permanent Magnet Synchronous Motor}
\acro{PWM}{Pulse Width Modulation}
\acro{SMO}{Sliding Mode Observer}
\acro{SR}{Synchronous Rectifier}
\acro{SVPWM}{Space Vector Pulse Width Modulation}
\acro{SWT}{Small Wind Turbine}
\acro{TM}{Torque Meter}
\acro{TSR}{Tip Speed Ratio}
\acro{VFD}{Variable Frequency Drive}
\acro{WECS}{Wind Energy Conversion System}


\acro{WTE}{Wind Turbine Emulator}
		
\end{acronym}